\documentclass{aa}  

\usepackage{graphicx}
\usepackage{natbib}
\bibpunct{(}{)}{;}{a}{}{,}

\usepackage{upgreek}
\usepackage{amsmath}
\usepackage{amssymb}
\usepackage{txfonts}
\usepackage[linkcolor=blue, bookmarks=true, bookmarksopen=true,
a4paper=true, citecolor=blue, urlcolor=blue, colorlinks=true,
pagecolor=blue, breaklinks=true, plainpages=false]{hyperref}
\usepackage[all]{hypcap}

\newcommand{\mdegr}{^{\circ}}

\newcommand{\earth}{\mathrm{\oplus}}
\newcommand{\sub}[1]{_\mathrm{#1}}
\newcommand{\dif}{\mathrm{d}}
\mathchardef\mhyphen="2D

\begin{document}

\title{Dusty tails of evaporating exoplanets}
\subtitle{I. Constraints on the dust composition}

  \author{
  R. van Lieshout\inst{1}
  \and
  M. Min\inst{1}
  \and
  C. Dominik\inst{1,2}
  }

  \institute{
  Anton Pannekoek Institute for Astronomy, University of Amsterdam, Science Park 904, 1098 XH Amsterdam, The Netherlands
  \\ \email{r.vanlieshout@uva.nl}
  \and
  Department of Astrophysics/IMAPP, Radboud University Nijmegen, P.O. Box 9010, 6500 GL Nijmegen, The Netherlands
  }

  \abstract
    {
    Recently, two exoplanet candidates have been discovered, KIC~12557548b and \mbox{KOI-2700b},
    whose transit profiles show evidence of a comet-like tail of dust trailing the planet,
    thought to be fed by the evaporation of the planet's surface.
    }
    {
    We aim to put constraints on the composition of
    the dust ejected by these objects
    from the shape of their transit light curves.
    }
    {
    We derive a semi-analytical expression for
    the attenuation of the dust cross-section in the tail,
    incorporating the sublimation of dust grains as well as their drift away from the planet.
    This expression shows that the length of the tail is
    highly sensitive to the sublimation properties of the dust material.
    We compute tail lengths
    for several possible dust compositions,
    and compare these to observational estimates of the tail lengths of KIC~12557548b and \mbox{KOI-2700b},
    inferred from their light curves.
    }
    {
    The observed tail lengths are consistent with dust grains composed of corundum (Al$_2$O$_3$)
    or iron-rich silicate minerals (e.g., fayalite, Fe$_2$SiO$_4$).
    Pure iron and carbonaceous compositions are not favoured.
    In addition, we estimate dust mass loss rates of
    $ 1.7 \pm 0.5 \mathrm{~M\sub{\earth}~Gyr^{-1} } $ for KIC~12557548b,
    and $ > 0.007 \mathrm{~M\sub{\earth}~Gyr^{-1} } $ ($1 \sigma $ lower limit) for \mbox{KOI-2700b}.
    }
    {}

 \keywords{ eclipses -- planets and satellites: composition
   -- planets and satellites: individual: (KIC~12557548b, \mbox{KOI-2700b}) }

  \maketitle

\section{Introduction}

The recently discovered exoplanet candidates \object{KIC~12557548}b (hereafter KIC~1255b) and \mbox{\object{KOI-2700}b}
show asymmetric transit profiles,
that can be explained by the occultation of stars by comet-like tails of dust
emanating from evaporating exoplanets \citep{2012ApJ...752....1R,2014ApJ...784...40R}.
The plausibility of this scenario has been strengthened by
quantitative modelling of the transit light curve
\citep{2012A&A...545L...5B,2013A&A...557A..72B,2014A&A...561A...3V},
and of the Parker-type thermal wind that can eject dust grains
from the planetary atmosphere \citep{2013MNRAS.433.2294P}.
Table~\ref{tbl:sys} lists the basic properties of the two systems.

If the evaporating-planet scenario is correct,
KIC~1255b and \mbox{KOI-2700b} may provide a rare chance to probe the interiors of small exoplanets.
Since the dust tails are thought to originate in the atmospheres of the planets,
which are fed by the evaporating planetary surfaces,
knowledge of the dust composition would
provide information about the composition of the planets.
This, in turn, would
shed light on the origin and evolutionary history of the planets, and would therefore
comprise
a valuable constraint
for theories of planet formation and evolution.

In this paper, we demonstrate that measurements of the length of a dust tail
can be used to put constraints on the
composition of the dust grains in the tail \citep[see also][]{2002Icar..159..529K}.
In Sect.~\ref{s:theory}, we show how the tail length is related to the rate at which dust grains
become smaller as a result of sublimation
and the rate at which they drift away from the planet.
In Sect.~\ref{s:applic}, we apply this theory
to the two candidate evaporating planets discovered thus far,
testing how well several possible dust species explain the observations.
Finally, in Sect.~\ref{s:discuss} we discuss our findings,
and in Sect.~\ref{s:conclusions} we list our conclusions.

\section{Cross-section decay in a dusty tail}
\label{s:theory}

In this section, we derive an expression for the decay of
extinction cross-section per unit angle $ W $
with angular separation from the planet $ \Delta\theta $.
This equation will describe
how the depth of the transit
depends on the mass loss rate of the planet,
and
how the length of the tail
depends on the material properties of the dust.
In our derivation we make use of the following assumptions,
whose validity is discussed in the derivation and in Sect.~\ref{s:assump}:
\begin{enumerate}
   \item \label{as:steady}
   The dust tail can be treated as if it were in steady state.
  \item \label{as:thin}
  The tail is radially optically thin throughout.
  \item \label{as:size}
  The cross-section in the tail is dominated by dust grains with the same initial size.
  \item \label{as:sphere}
  The grains can be treated as simple spheres.
  \item \label{as:qs}
  The optical efficiencies of the grains (i.e., $ Q\sub{abs} $, $ Q\sub{scat} $, etc.)
  do not change significantly as the grains become smaller.
  \item \label{as:t_subl}
  The dust grains survive for at least one orbit.
  \item \label{as:dsdt}
  The orbit-averaged sublimation rate of a dust grain does not change substantially as a grain becomes smaller.
  \item \label{as:gas}
  Sublimation of dust grains can be treated as in vacuum.
  Gas released by the planet does not offset the sublimation through recondensation.
  \item \label{as:ecc_p}
  The planet's orbit is circular.
  \item \label{as:r_p}
  The size of the planet is negligible compared to the length of the dust tail
  (i.e., grains are released from a single point).
  \item \label{as:temp}
  Dust temperatures are determined by absorption and reradiation of stellar radiation.
  \item \label{as:dyn}
  The dynamics of the dust grains are dominated by drift due to radiation pressure.
  The initial relative velocities with which the particles are released from the planet are negligible.
\end{enumerate}

\begin{table}
  \centering
  \caption{Host star and system parameters of the two evaporating exoplanet candidates}
  \label{tbl:sys}
  {
  \renewcommand{\arraystretch}{1.16}
  \begin{tabular}{lccc}
  \hline
  & KIC~1255b & KOI-2700b & Refs. \\
  \hline
  $ T\sub{eff,\star} $ [K] & $ 4550\substack{+140 \\ -131} $ & $ 4296\substack{+131 \\ -146} $ & H14 \\
  $ M\sub{\star} $ [M$ \sub{\odot} $] & $ 0.666\substack{+0.067 \\ -0.059} $ & $ 0.546 \pm 0.044 $ & H14 \\
  $ R\sub{\star} $ [R$ \sub{\odot} $] & $ 0.660\substack{+0.060 \\ -0.059} $ & $ 0.540\substack{+0.048 \\ -0.051} $ & H14 \\
  $ L\sub{\star} $ [L$ \sub{\odot} $] & $ 0.168\substack{+0.037 \\ -0.036} $ & $ 0.089\substack{+0.019 \\ -0.021} $ & \\
  \hline
  $ P\sub{p} $ [days] & $ 0.6535538(1) $ & $ 0.910022(5) $ & V14, R14 \\
  $ a\sub{p} $ [AU] & $ 0.0129(4) $ & $ 0.0150(4) $ & \\
  \hline
  \end{tabular}
  \tablefoot{
  From top to bottom:
  stellar effective temperature,
  stellar mass,
  stellar radius,
  stellar luminosity (derived from $ T\sub{eff,\star} $ and $ R\sub{\star} $),
  orbital period of the candidate planet,
  and corresponding semi-major axis (as given by Kepler's third law).
  Numbers in brackets indicate the uncertainty on the last digit.
  }
  \tablebib{
  H14~\citet{2014ApJS..211....2H};
  R14~\citet{2014ApJ...784...40R};
  V14~\citet{2014A&A...561A...3V}.
  }
  }
\end{table}

\begin{figure*}
  \includegraphics[width=\linewidth]{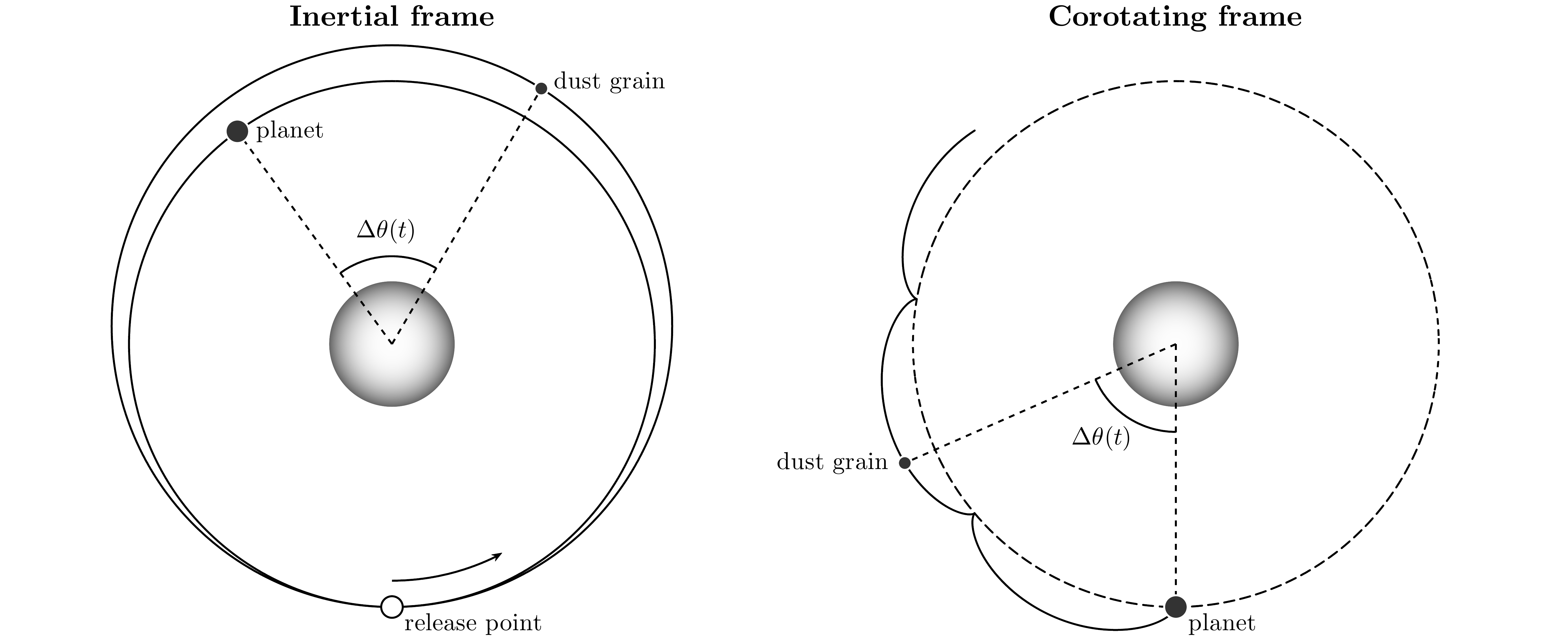}
  \caption{
  Diagrams of the path followed by a dust grain released from an evaporating planet (ignoring sublimation of the dust grain)
  for 3~orbital periods of the planet, in the inertial frame and in the frame corotating with the planet.
  The positions of the planet and the dust particle are indicated
  at a time 1.6 orbital periods of the planet after release.
  }
  \label{fig:sketch}
\end{figure*}

The basic geometry of the system is shown in Fig.~\ref{fig:sketch}.
We express the collective extinction cross-section of the dust
as a fraction of the area of the stellar disk.
Its angular density $ W $ therefore has units [rad$^{-1}$].
Under the assumptions listed above,
$ W( \Delta\theta ) $ can be written as
\begin{equation}
  \label{eq:tail_def1}
  W( \Delta\theta )
    = \frac{ \dif n }{ \dif \Delta\theta }
      \frac{ \sigma\sub{ext}( \Delta\theta ) }{ \pi R\sub{\star}^2 },
\end{equation}
where $ \dif n / \dif \Delta\theta $ is the number of particles per unit angle,
$ \sigma\sub{ext} $ is the extinction cross-section of a single dust grain,
and $ R\sub{\star} $ is the stellar radius.
The angular number density can be expanded into
\begin{equation}
  \label{eq:tail_def2}
  \frac{ \dif n }{ \dif \Delta\theta }
    = \frac{ \dif n }{ \dif t }
      \left< \frac{ \dif t }{ \dif \Delta\theta } \right>
    = \frac{ \dot{M}\sub{d} }{ m\sub{0} }
      \left< \frac{ \dif t }{ \dif \Delta\theta } \right>.
\end{equation}
Here,
$ t $ denotes time,
$ \dif n / \dif t $ is the rate at which the planet releases particles,
the brackets $ \left< \, \dots \right> $ indicate averaging over the orbit of the dust grain,
$ \dot{M}\sub{d} $ is
the dust mass loss rate of the planet (i.e., excluding the mass lost in gas),
and $ m\sub{0} $ is the mass of an individual dust grain when it leaves the planet.
The extinction cross-section of a single dust grain as a function of angular separation $ \sigma\sub{ext}( \Delta\theta ) $
can be inferred
from its derivative, which
can be expanded into
\begin{equation}
  \label{eq:tail_def3}
  \frac{ \dif \sigma\sub{ext} }{ \dif \Delta\theta }
    = \frac{ \dif \sigma\sub{ext} }{ \dif s }
      \left< \frac{ \dif s }{ \dif t } \right>
      \left< \frac{ \dif t }{ \dif \Delta\theta } \right>.
\end{equation}
Here,
$ s $ is the grain radius.

In the following subsections,
we derive expressions for
the size dependence of extinction cross-section $ \dif \sigma\sub{ext} / \dif s $
(Sect.~\ref{s:sigma_ext}),
the azimuthal drift rate $ \left< \dif \Delta\theta / \dif t \right> $ (Sect.~\ref{s:omega_syn}),
and the grain radius change rate $ \left< \dif s / \dif t \right> $ (Sect.~\ref{s:subl}).
These are then combined using Eqs.~\eqref{eq:tail_def1}--\eqref{eq:tail_def3}
to yield the equation for the decay of extinction cross-section
(Sect.~\ref{s:analytic}).

\subsection{Extinction cross-section}
\label{s:sigma_ext}

\begin{figure}
  \includegraphics[width=\linewidth]{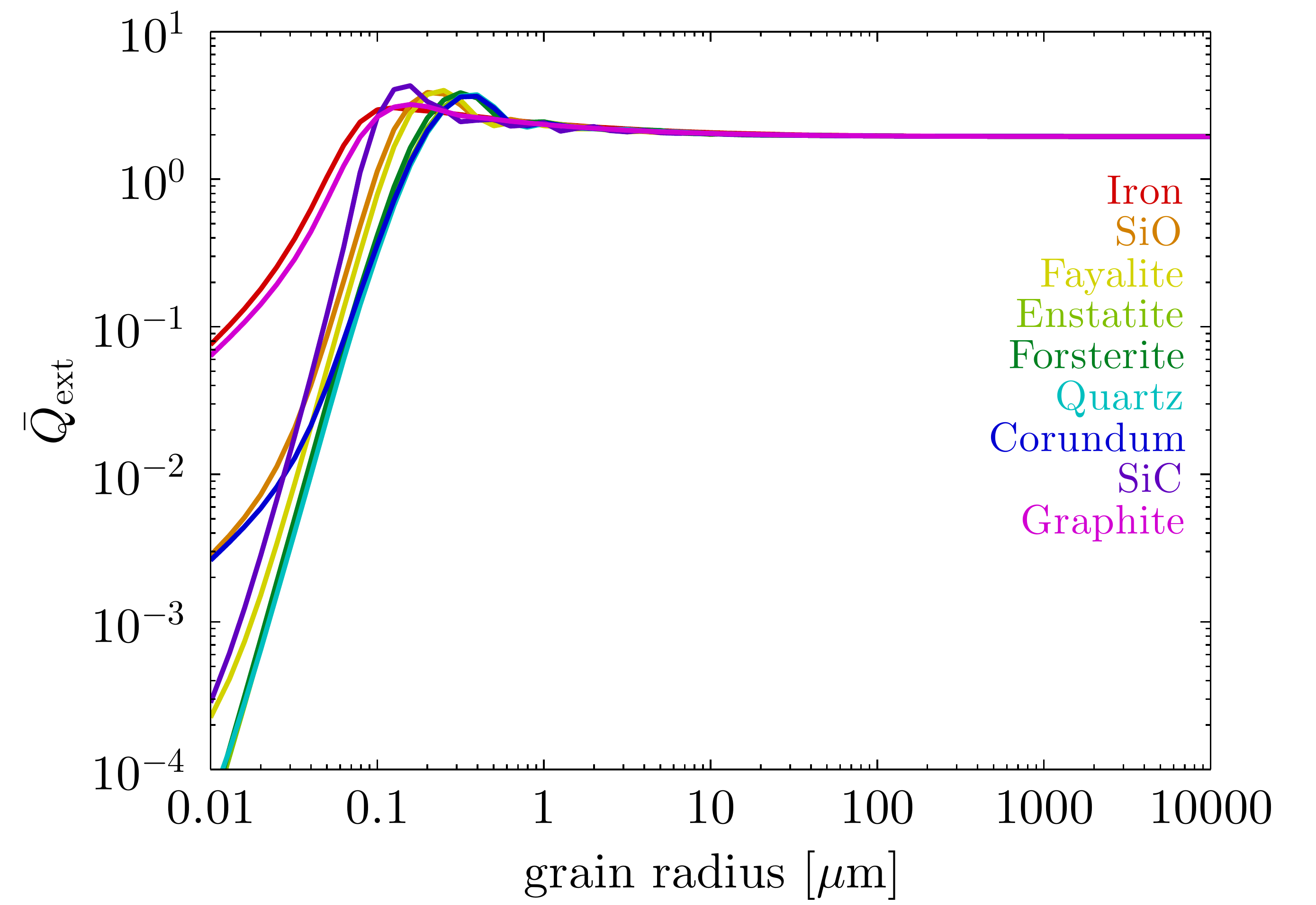}
  \caption{
  Wavelength-averaged extinction efficiency $ \bar{Q}\sub{ext} $ as a function of grain radius $ s $
  for the dust species considered in this study \citep[cf. Fig.~13 of][]{2014ApJ...786..100C}.
  Since KIC~12557548 and \mbox{KOI-2700} are of similar stellar type,
  these values apply to both stars.
  }
  \label{fig:mat_q_ext}
\end{figure}

The extinction cross-section of a spherical dust grain with radius $ s $
can be expressed as
\begin{equation}
  \label{eq:sigma_ext}
  \sigma\sub{ext}( s ) = \pi \bar{Q}\sub{ext}( s ) s^2.
\end{equation}
Here, $ \bar{Q}\sub{ext} $ is the extinction efficiency
averaged over the stellar spectrum and the spectral response function of the instrument.
We calculate monochromatic extinction efficiencies
from (material dependent) refractive index data
using \citet{1908AnP...330..377M} theory.
The sources of refractive index data are listed in Table~\ref{tbl:opt}.
For the stellar spectra we take \citet{1993KurCD..13.....K} models.\footnote{\label{fn:f_star}
For both stars, we use model atmospheres with
an effective temperature of $ T\sub{eff,\star} = 4500~\mathrm{K} $ and
a surface gravity of $ \log g  = 4.5 $,
compatible with the stellar parameters given by \citet{2014ApJS..211....2H}.
}
The \textit{Kepler} response function is
described by \citet{2010ApJ...713L..79K}.
Figure~\ref{fig:mat_q_ext} shows the resulting extinction efficiencies.

\begin{table}[!t]
  \centering
  \caption{Bulk densities and sources for optical properties of the dust species considered in this study}
  \label{tbl:opt}
  \begin{tabular}{lr@{.}lc}
  \hline
  Dust species & \multicolumn{2}{c}{$ \rho\sub{d} $} & Refs. for opt. prop. \\
  & \multicolumn{2}{c}{[g cm$ ^{-3} $]} & \\
  \hline
  Iron (Fe) & 7 & 87 & O88 \\
  Silicon monoxide (SiO) & 2 & 13 & P85, W13 \\
  Cryst. fayalite (Fe$_2$SiO$_4$) & 4 & 39 & Z11, F01 \\
  Cryst.\tablefootmark{a} enstatite (MgSiO$_3$) & 3 & 20 & D95, J98 \\
  Cryst. forsterite (Mg$_2$SiO$_4$) & 3 & 27 & Z11, F01 \\
  Quartz (SiO$_2$) & 2 & 60 & Z13 \\
  Corundum (Al$_2$O$_3$) & 4 & 00 & K95 \\
  Silicon carbide (SiC) & 3 & 22 & L93 \\
  Graphite (C) & ~~~~2 & 16 & D84 \\
  \hline
  \end{tabular}
  \tablefoot{
  \tablefoottext{a}{The optical properties at wavelengths below 8~$\upmu$m use amorphous enstatite.}
  }
  \tablebib{
  D84~\citet{1984ApJ...285...89D};
  D95~\citet{1995A&A...300..503D};
  F01~\citet{2001A&A...378..228F};
  J98~\citet{1998A&A...339..904J};
  K95~\citet{1995Icar..114..203K};
  L93~\citet{1993ApJ...402..441L};
  O88~\citet{Ordal:88};
  P85~\citet{1985hocs.book.....P};
  W13~\citet{2013A&A...553A..92W};
  Z11~\citet{2011A&A...526A..68Z};
  Z13~\citet{2013A&A...553A..81Z}.
  }
\end{table}

Figure~\ref{fig:mat_q_ext} shows
that the extinction efficiency does not change substantially with grain size
($ \dif \bar{Q}\sub{ext} / \dif s \approx 0 $)
for $ s \gtrsim 0.1$--0.5~$\upmu$m, depending on material type.
In this regime,
we can make the approximation
\begin{equation}
  \label{eq:dsigma_extds}
  \frac{ \dif \sigma\sub{ext} }{ \dif s }
    = 2 \pi \bar{Q}\sub{ext} s.
\end{equation}

\subsection{Drift due to radiation pressure}
\label{s:omega_syn}

\begin{figure}
  \includegraphics[width=\linewidth]{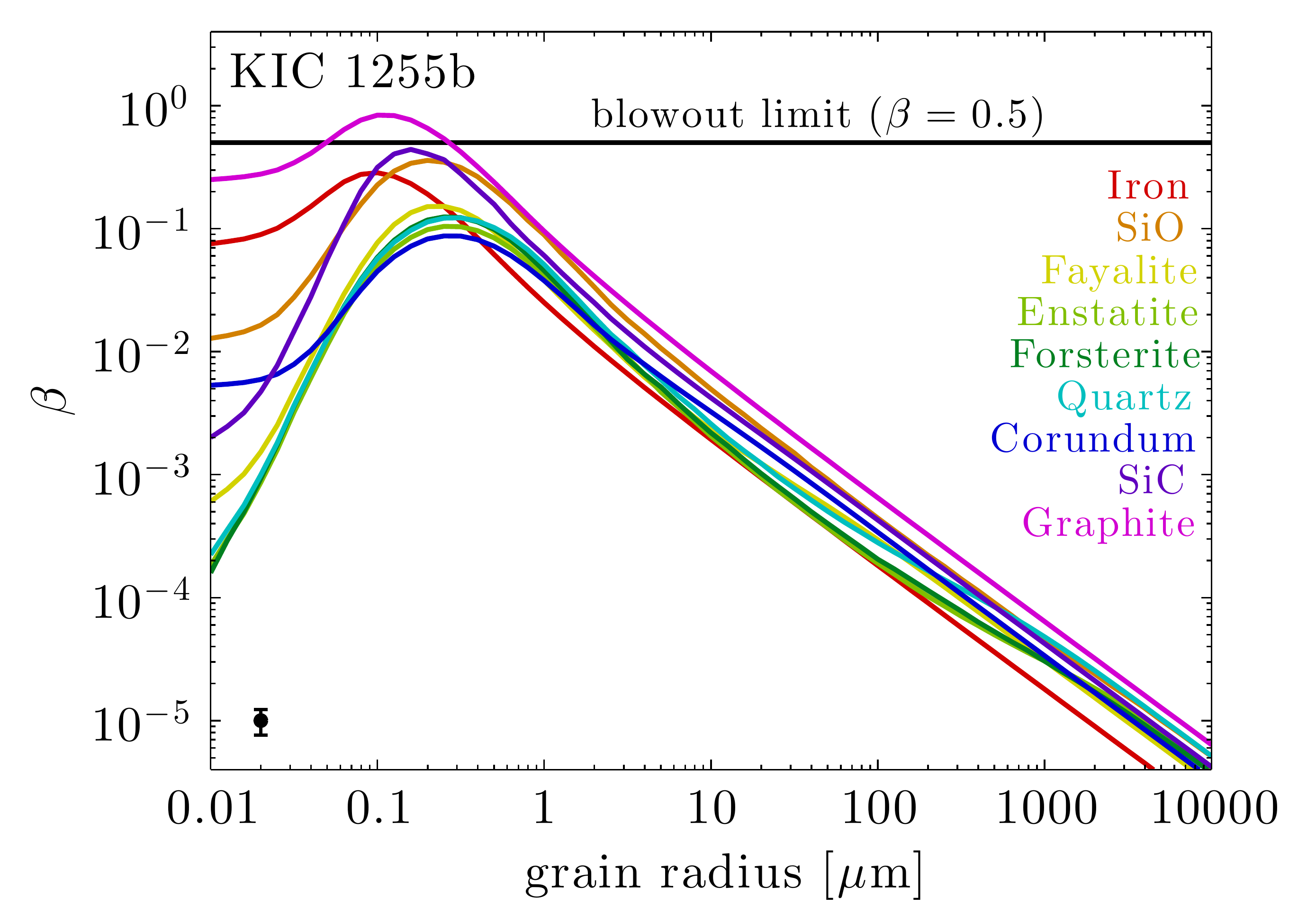}
  \caption{
  Radiation pressure to gravitational force ratio $ \beta $
  as a function of grain radius $ s $
  for KIC~1255b.
  Different colours correspond to the materials listed in Table~\ref{tbl:opt}.
  Uncertainties (due to uncertainties in stellar parameters) are indicated by the error bar
  in the bottom left.
  The same figure for \mbox{KOI-2700b} is qualitatively the same,
  but the \mbox{$ \beta $ values} are lower by a factor of about~1.5.
  }
  \label{fig:mat_beta}
\end{figure}

The dynamics of small dust grains is significantly affected by radiation pressure from the star.
Since the radiation pressure force scales the same way with distance from the star as gravity,
it is parametrised by $ \beta $, the ratio between the norms of the direct radiation pressure force and the gravitational force
(i.e., $ \beta = | F\sub{rad} / F\sub{grav} | $).
For spherical dust grains,
this parameter is given by \citep[e.g.,][]{1979Icar...40....1B}
\begin{equation}
  \label{eq:beta}
  \beta = \frac{ 3 }{ 16 \pi c G }
    \frac{ L\sub{\star} }{ M\sub{\star} }
    \frac{ \bar{Q}\sub{pr}( s ) }{ \rho\sub{d} s }.
\end{equation}
Here,
$ c $ is the speed of light,
$ G $ is the gravitational constant,
$L\sub{\star}$ is the stellar luminosity,
$ M\sub{\star} $ is the stellar mass,
$ \bar{Q}\sub{pr} $ is the radiation pressure efficiency averaged over the stellar spectrum,
and $ \rho\sub{d} $ is the bulk density of the dust.
Figure~\ref{fig:mat_beta} shows the \mbox{$ \beta $ ratios}
for dust grains of different compositions,
using the stellar parameters listed in Table~\ref{tbl:sys},
bulk densities from Table~\ref{tbl:opt},
and \mbox{$ \bar{Q}\sub{pr} $ values} derived from the Mie calculations described
in Sect.~\ref{s:sigma_ext}.

After being released from a planet that follows a circular Keplerian orbit,
radiation pressure-affected particles will
continue on new Keplerian orbits
with eccentricities $ e\sub{d} $, semi-major axes $ a\sub{d} $, and periods $ P\sub{d} $ given by
\citep[e.g.,][]{2014ApJ...784...40R}
\begin{equation}
  \label{eq:ecc_beta}
  e\sub{d} = \frac{ \beta }{  1 - \beta };
  \qquad
  a\sub{d} = a\sub{p} \frac{  1 - \beta }{ 1 - 2 \beta };
  \qquad
  P\sub{d} = P\sub{p} \frac{  1 - \beta }{ ( 1 - 2 \beta )^{3/2} }.
\end{equation}
Here, $ a\sub{p} $, and $ P\sub{p} $
denote the semi-major axis and orbital period of the planet, respectively.
We note that particles with $ \beta \geq 0.5 $
leave the system on unbound orbits.

Because of its eccentric orbit, a dust grain will drift away azimuthally from the planet
and oscillate in radial distance, giving rise to a rosette-like path in the corotating frame
(see Fig.~\ref{fig:sketch}).
Averaged over the particle's orbit
(but ignoring sublimation),
the azimuthal drift rate equals the synodic orbital frequency,
which is given by \citep{2014ApJ...784...40R}
\begin{equation}
  \label{eq:omega_syn}
  \omega\sub{syn}
    = \frac{ 2 \pi  }{ P\sub{p} } - \frac{ 2 \pi  }{ P\sub{d} } 
    = \frac{ 2 \pi  }{ P\sub{p} } \frac{ 1 - \beta - ( 1 - 2 \beta )^{3/2} }{ 1 - \beta }
    \approx \frac{ 4 \pi \beta }{ P\sub{p} }.
\end{equation}
The mathematical approximation in the last step
gives an error of
less than about $ 5\%$ for all $\beta < 0.5$,
and
less than $0.5\%$ for all $\beta < 0.1$.
Compared to the actual (non-orbit-averaged) drift,
this constant drift rate gives errors in $ \Delta\theta $ of less than $15\%$ after the first orbit
for $ \beta < 0.1 $ (see also Fig.~\ref{fig:cusp_th}).

The azimuthal drift of a grain is described by
$ \Delta\theta( t ) = \omega\sub{syn} t $.
Since $ \omega\sub{syn} $ depends on $ \beta $,
a particle's drift rate will change as it becomes smaller because of sublimation.
Using
the assumption that the radiation pressure efficiency
does not change substantially with decreasing grain size
($ \dif \bar{Q}\sub{pr} / \dif s \approx 0 $,
which roughly holds for $ s \gtrsim 0.1$--0.5~$\upmu$m, depending on material type),
we find
\begin{align}
  \left< \frac{ \dif \Delta\theta }{ \dif t } \right>
    & = \omega\sub{syn}
      + \frac{ \dif \omega\sub{syn} }{ \dif \beta }
        \frac{ \dif \beta }{ \dif s }
        \left< \frac{ \dif s }{ \dif t } \right> t \\
  \label{eq:drift_rate}
    & \approx \frac{ 4 \pi \beta }{ P\sub{p} }
      \left( 1 - \frac{ t }{ s } \left< \frac{ \dif s }{ \dif t } \right> \right)
    = \frac{ 4 \pi \beta }{ P\sub{p} } \frac{ s\sub{0} }{ s },
\end{align}
where $ s\sub{0} $ denotes the initial grain radius.
The final step uses
the assumption that the orbit-averaged sublimation rate remains constant,
which implies that
the evolution of a grain's radius is described by
$ s(t) = s_0 + \left< \dif s / \dif t \right> t $.

\subsection{Dust sublimation}
\label{s:subl}

\begin{table*}
 \centering
 \caption{Sublimation characteristics of the dust species considered in this study}
 \label{tbl:subl}
 \begin{tabular}{lr@{.}lr@{.}lr@{ $\pm$ }lr@{ $\pm$ }lr@{ -- }lcc}
 \hline
 Dust species & \multicolumn{2}{c}{ $ \; \; \mu $ } & \multicolumn{2}{c}{ $ \alpha \; \; $ }
   & \multicolumn{2}{c}{ $ \mathcal{A} $ } & \multicolumn{2}{c}{ $ \mathcal{B} $ }
   & \multicolumn{2}{c}{ Temp. range\tablefootmark{e}  } & Refs. & Notes \\
  & \multicolumn{2}{c}{} & \multicolumn{2}{c}{} & \multicolumn{2}{c}{ [K] } & \multicolumn{2}{c}{} & \multicolumn{2}{c}{ [K]~~~ } & & \\
 \hline
 Iron (Fe) & 55 & 845 & 1 & 0 & 48354 & 1151 & 29.2 & 0.7 & 1573 & 1973 & F04 & \\
 Silicon monoxide (SiO) & 44 & 085 & 0 & 04 & 49520 & 1400 & 32.5 & 1.0 & 1275 & 1525 & G13 & \\
 Cryst. fayalite (Fe$_2$SiO$_4$) & 203 & 774 & 0 & 1\tablefootmark{a} & 60377 & 1082 & 37.7 & 0.7 & 1373 & 1433 & N94 & \\
 Cryst. enstatite (MgSiO$_3$) & 100 & 389 & 0 & 1\tablefootmark{a} & 68908 & 8773 & 38.1 & 5.0 & 1573 & 1923 & M88, K91 & \\
 Cryst. forsterite (Mg$_2$SiO$_4$) & 140 & 694 & 0 & 1\tablefootmark{b} & 65308 & 3969 & 34.1 & 2.5 & 1673 & 2133 & N94 & \\
 Quartz (SiO$_2$) & 60 & 084 & 1 & 0 & 69444 & 3447 & 33.1 & 1.8 & 1833 & 1958 & H90 & (f) \\
 Corundum (Al$_2$O$_3$) & 101 & 961 & 0 & 1\tablefootmark{c} & 77365 & 3868\tablefootmark{d} & 39.3 & 2.0\tablefootmark{d} & \multicolumn{2}{c}{} & L08 & \\
 Silicon carbide (SiC) & 40 & 10 & 0 & 1\tablefootmark{a} & 78462 & 3923\tablefootmark{d} & 37.8 & 1.9\tablefootmark{d} & 1500 & 2000 & L93 & (g) \\
 Graphite (C) & 12 & 011 & 0 & 1\tablefootmark{a} & 93646 & 503 & 36.7 & 1.8\tablefootmark{d} & 2400 & 3000 & Z73 & (h) \\
 \hline
 \end{tabular}
 \tablefoot{
 \tablefoottext{a}{For materials for which no measuments of the evaporation coefficient are available
 we arbitrarily adopt $ \alpha = 0.1 $.}
 \tablefoottext{b}{From \citet{2010LNP...815...61G}.}
 \tablefoottext{c}{From \citet{2004Icar..169..216S}.}
 \tablefoottext{d}{Where no uncertainty on sublimation parameters was reported,
 we arbitrarily set the standard deviation to 5\%.}
 \tablefoottext{e}{Range of temperatures for which $ \mathcal{A} $ and $ \mathcal{B} $ were determined.}
 \tablefoottext{f}{These sublimation parameters were measured using a different polymorph of SiO$_2$, namely high cristobalite.
 Furthermore, they assume the sublimation proceeds following the reaction SiO$_2$(solid) $ \rightarrow $ SiO$_2$(gas),
 for which the evaporation coefficient $ \alpha $ is close to unity \citep{1990Natur.347...53H}.}
 \tablefoottext{g}{Using the equilibrium vapour pressure of SiC$_2$.}
 \tablefoottext{h}{At the temperatures relevant to this study,
 graphite sublimates mostly as \mbox{C$_3$-clusters}, and the quoted parameters correspond to this component \citep{zavitsanos:2966}.}
 }
  \tablebib{
  F04~\citet{ferguson2004};
  G13~\citet{2013A&A...555A.119G};
  H90~\citet{1990Natur.347...53H};
  K91~\citet{kushiro91};
  L93~\citet{Lilov199365};
  L08~\citet{Lihrmann2008649};
  M88~\citet{1988AmMin..73....1M};
  N94~\citet{1994GeCoA..58.1951N};
  Z73~\citet{zavitsanos:2966}.
  }
\end{table*}

While a dust particle drifts away from the planet,
it will gradually sublimate
because of the equilibrium temperature it reaches when illuminated by the stellar radiation.
Since we consider spherical dust grain,
whose mass and surface area are given by
$ m = 4/3 \pi s^3 \rho\sub{d} $ and $ A = 4 \pi s^2 $, respectively,
the orbit-averaged rate at which the grain radius changes can be rewritten as
\begin{equation}
  \label{eq:dsdt}
  \left< \frac{ \dif s }{ \dif t } \right>
    = \frac{ \dif s }{ \dif m } \left< \frac{ \dif m }{ \dif t } \right>
    = - \frac{ \dif s }{ \dif m } A \left< J \right>
    = - \frac{ \left< J \right> }{ \rho\sub{d} }.
\end{equation}
Here, $ J $ denotes the mass loss flux
from the surface of the dust grain
(units: [g~cm$^{-2}$~s$^{-1}$]; positive for mass loss).

According to the kinetic theory of gases,
a solid surface with temperature $ T $ in vacuum
loses mass at a rate of
\citep{langmuir1913}
\begin{equation}
  \label{eq:subl}
  J( T ) = \alpha p\sub{v}( T )
    \sqrt{ \frac{ \mu m\sub{u} }{ 2 \pi k\sub{B} T } }.
\end{equation}
Here,
$ \alpha $ is the evaporation coefficient
that parameterises kinetic inhibition of the sublimation process
(which we assume to be independent of temperature),
$ p\sub{v} $ is the material-specific phase-equilibrium vapour pressure,
$ \mu $ is the molecular weight of the molecules that sublimate,\footnote{
More precisely, $ \mu $ should reflect the average molecular weight
of the molecules recondensing from the gas phase
in an equilibrium between sublimation and condensation.
For simplicity, we use the molecular weights of the dust compound.
Discrepancies may result in errors of the order of unity in the final value of $ J $,
which are negligible compared to the uncertainties on $ \alpha $ and $ p\sub{v} $.}
$ m\sub{u} $ is the atomic mass unit,
and $ k\sub{B} $ is the Boltzmann constant.
The temperature dependence of the equilibrium vapour pressure
can be approximated by
\begin{equation}
  \label{eq:pres_vap2}
  p\sub{v} = \exp ( \, - \mathcal{A} / T + \mathcal{B} \, ) \; \mathrm{ dyn~cm^{-2} },
\end{equation}
where $ \mathcal{A} $ and $ \mathcal{B} $ are material-dependent sublimation parameters
that can be determined
experimentally.
Table~\ref{tbl:subl} lists estimates of the sublimation parameters for the materials we consider.

A particle on an eccentric orbit has different temperatures $ T\sub{d} $
at different parts of its orbit.
When computing the orbit-averaged mass loss flux,
$ J( T\sub{d} ) $ has to be weighed by the time spent in each part of the orbit.
This gives
\begin{equation}
  \label{eq:subl_mean}
  \left< J \right>
    = \frac{ 1 }{ 2 \pi } \int \limits_{ 0 }^{ 2 \pi }
      J [ T\sub{d} ( \theta\sub{d} ) ]
      \frac{ ( 1 - e\sub{d}^2 )^{3/2} }{ (1 + e\sub{d} \cos \theta\sub{d} )^2 } \, \dif \theta\sub{d},
\end{equation}
where $ \theta\sub{d} $ is the true anomaly of the dust grain,
and $ T\sub{d} $ depends on $ \theta\sub{d} $ through distance
$ r = a\sub{d} ( 1 - e\sub{d}^2 ) / ( 1 + e\sub{d} \cos \theta\sub{d} ) $.

\begin{figure*}
  \includegraphics[width=\linewidth]{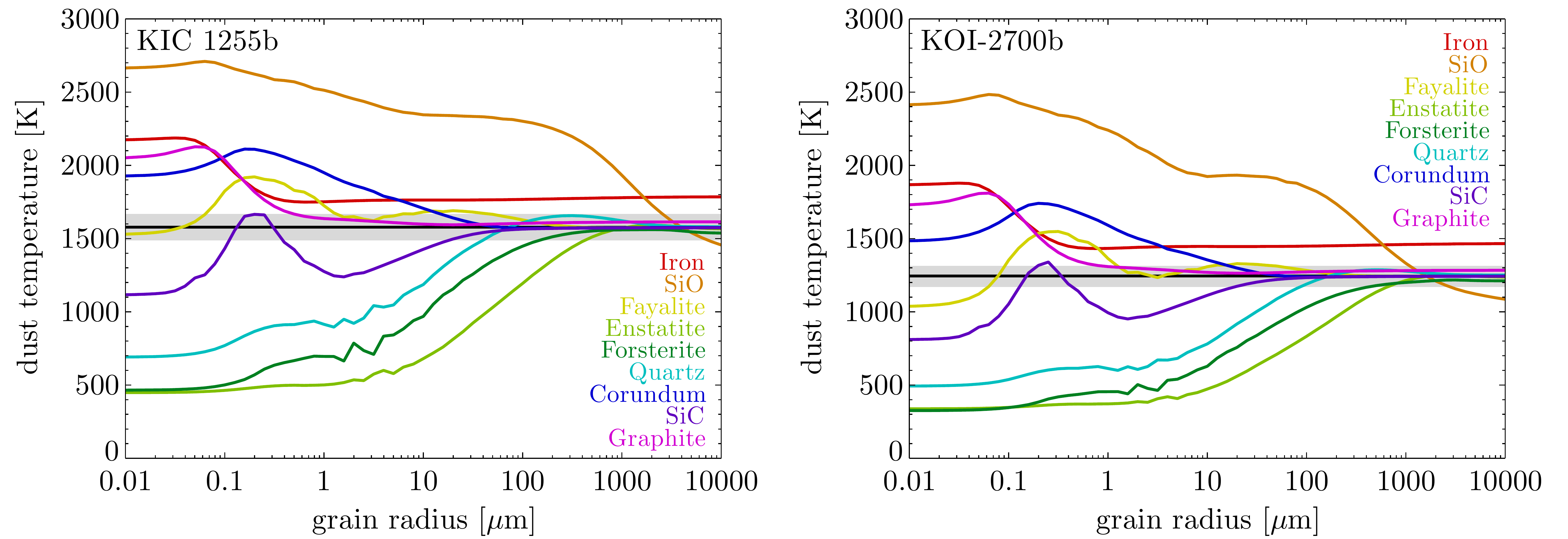}
  \caption{
  Dust temperatures $ T\sub{d} $ at the orbital distance of the planet as a function of grain radius $ s $
  for various dust materials.
  The horizontal black lines mark
  the black-body temperatures at the distances of the planets.
  The grey bands indicate the uncertainties on the black-body temperatures,
  and can be used as estimates of the typical uncertainties on the grain temperatures.
  }
  \label{fig:mat_temp}
\end{figure*}

The temperature $ T\sub{d}( s, r ) $ of a dust grain with size $ s $ at distance $ r $
 from the (centre of the) star
can be found by solving the energy balance
\begin{equation}
  \label{eq:temp_d}
  \frac{ \Omega( r ) }{ \pi } \! \! \int \! Q\sub{abs}( s, \lambda ) \, F\sub{ \star }( \lambda ) \, \dif \lambda
    = 4  \! \int \! Q\sub{abs}( s, \lambda ) \, B_\lambda( \lambda, T\sub{d} ) \, \dif \lambda.
\end{equation}
Here,
$ \lambda $ denotes wavelength,
$ Q\sub{abs} $ is the monochromatic absorption efficiency of the dust grain,
$ F\sub{ \star } $ is the stellar spectrum,
$ B_\lambda $ denotes the Planck function,
and $ \Omega( r ) $ is the solid angle subtended by the star as seen from a distance $ r $,
which is given by
$ \Omega( r ) = 2 \pi \left[ 1 - \sqrt{ 1 - ( R\sub{\star} / r )^2 } \, \right] $.
Figure~\ref{fig:mat_temp} shows the grain temperatures
of dust particles made of several different materials
at the orbital distance of KIC~1255b and \mbox{KOI-2700b}.
We use \mbox{$ Q\sub{abs} $ values} derived from the Mie calculations described in Sect.~\ref{s:sigma_ext}.
For $ F\sub{ \star }( \lambda ) $, we use \citet{1993KurCD..13.....K} models (see footnote~\ref{fn:f_star})
scaled with the stellar luminosity.
At large grain sizes,
temperatures of most of the investigated materials
approach the black-body temperature
\begin{equation}
  \label{eq:temp_bb}
  T\sub{bb}(r) = \left[ \frac{ \Omega( r ) }{ 4 \pi } \right]^{1/4} T\sub{ eff, \star }.
\end{equation}
At the distance of the planet, this gives
$ T\sub{bb}( a\sub{p} ) = 1577\substack{+91 \\ -89} $~K for KIC~1255b
and $ T\sub{bb}( a\sub{p} ) = 1244\substack{+69 \\ -75} $~K for \mbox{KOI-2700b}
(indicated by the horizontal bands in Fig.~\ref{fig:mat_temp}).

\subsection{The decay equation}
\label{s:analytic}

We can now combine the above equations into
an expression for the decay of extinction cross-section.
By combining
Eqs.~\eqref{eq:tail_def3}--\eqref{eq:beta},
\eqref{eq:drift_rate}, and \eqref{eq:dsdt},
we find
\begin{equation}
  \label{eq:tail_decay2}
  \frac{ \dif \sigma\sub{ext} }{ \dif \Delta\theta }
    \approx - \frac{ 8 c G }{ 3 }
      \frac{ M\sub{\star} P\sub{p} }{ L\sub{\star} }
      \frac{ \left< J \right> }{ \bar{Q}\sub{pr}( s ) }
      \sqrt{ \frac{ \sigma\sub{ext}^3 }{ \sigma\sub{ext,0} } },
\end{equation}
where $ \sigma\sub{ext,0} $ is a particle's extinction cross-section when it is released from the planet.
Integration
then yields
\begin{equation}
  \label{eq:tail_decay_sig}
  \sigma\sub{ext}( \Delta\theta )
    \approx \sigma\sub{ext,0}
    \left( 1 + \frac{ \Delta\theta }{ 4 \Delta\theta\sub{tail} } \right)^{-2},
\end{equation}
where $ \Delta\theta\sub{tail} $ is the characteristic angle of the tail's decay (defined below).
Inserting this into Eq.~\eqref{eq:tail_def1}, together with Eqs.~\eqref{eq:tail_def2}, \eqref{eq:beta} and \eqref{eq:drift_rate}, gives
the final expression for the decay of extinction cross-section per unit angle with angular separation from the planet
\begin{subequations}
  \label{eq:tail_decay_total}
  \begin{align}
    \label{eq:tail_decay3}
    W( \Delta\theta )
      & \approx W\sub{0} \left( 1 + \frac{ \Delta\theta }{ 4 \Delta\theta\sub{tail} } \right)^{-4} \\
    \label{eq:w_0}
    W\sub{0}
      & = \frac{ c G }{ \pi }
      \frac{ M\sub{\star} P\sub{p} \dot{M}\sub{d} }{ L\sub{\star} R\sub{\star}^2 }
      \frac{ \bar{Q}\sub{ext}( s ) }{ \bar{Q}\sub{pr}( s ) } \\
    \label{eq:tail_theory}
    \Delta\theta\sub{tail}
      & = \frac{ 3 }{ 16 c G }
        \frac{ L\sub{\star} }{ M\sub{\star} P\sub{p} }
        \frac{ \bar{Q}\sub{pr}( s ) }{ \left< J \right> }.
  \end{align}
\end{subequations}

\section{Constraints from observed light curves}
\label{s:applic}

With the theoretical background in place,
we now investigate
what constraints can be derived from
the observed light curves of
KIC~1255b and \mbox{KOI-2700b}.
To do this properly,
one would have to use a transit model
that computes the light curve resulting from the occultation of a star by a dust tail
whose extinction cross-section distribution is given by Eq.~\eqref{eq:tail_decay3}.
Such models exist \citep{2012A&A...545L...5B,2013A&A...557A..72B,2014A&A...561A...3V,2014ApJ...784...40R},
but they generally adopt
an ad hoc exponential decay for $ W( \Delta\theta ) $.
While the decay predicted by Eq.~\eqref{eq:tail_decay3} is shallower than exponential,
the profiles are similar enough
to use the normalisation constants and \mbox{$ e $-folding} angles
derived from light curve fits as estimates for
$ W\sub{0} $ and $ \Delta\theta\sub{tail} $, respectively.\footnote{
The two equations are identical to first order at $ \Delta\theta = 0 $, and,
for a large exponent, Eq.~\eqref{eq:tail_decay3} approaches exponential decay
\citep[cf.][and see their Fig.~12]{2014ApJ...784...40R}:
\begin{equation*}
  \label{eq:exp_limit}
  \lim_{ n \rightarrow \infty } \left( 1 + \frac{ \Delta\theta }{ n \Delta\theta\sub{tail} } \right)^{-n}
    = \exp \left( \frac{ \Delta\theta }{ \Delta\theta\sub{tail} } \right).
\end{equation*}
}

The literature values we adopt for
the cross-section distribution parameters
are listed in Table~\ref{tbl:obs}.
\mbox{KOI-2700b} only has a lower limit on $ W\sub{0} $.
This is because its ingress is not resolved well,
which allows tails with higher cross-section densities
at higher transit impact parameters.
We determined the lower limit on $ W\sub{0} $
from the transit depth of $ \delta = 360 \pm 29 $~ppm \citep{2014ApJ...784...40R}.
In the limit that the tail is short compared to the chord on the stellar disk crossed by the planet,\footnote{
I.e., $ \Delta\theta\sub{tail} \ll \hat{r}\sub{c} $,
where $ \hat{r}\sub{c} $ is the angle subtended by the chord
\citep[see Fig.~4 of][]{2012A&A...545L...5B}.}
the transit depth corresponds to the total amount of extinction cross-section in the tail,
which
can be found by integrating Eq.~\eqref{eq:tail_decay_total}.
Hence, we find
\begin{equation}
  \label{eq:depth}
  \delta \le \int \limits_{ 0 }^{ \infty } W( \Delta\theta ) \dif \Delta\theta
    \approx \frac{ 4 }{ 3 } W\sub{0} \Delta\theta\sub{tail}
    = \frac{ 1 }{ 4 \pi }
    \frac{ \dot{M}\sub{d} }{ R\sub{\star}^2 }
    \frac{ \bar{Q}\sub{ext}( s ) }{ \left< J \right> }.
\end{equation}
This ignores the limb darkening of the star
as well as the forward scattering of starlight into the line of sight.
It also assumes that the planet
does not contribute significantly to the flux deficit.

\begin{table}
  \centering
  \caption{
  Constraints on the
  dust tail cross-section distribution parameters
  derived from transit light curve fitting
  }
  \label{tbl:obs}
  \begin{tabular}{lcc}
  \hline
  & KIC~1255b & KOI-2700b \\
  \hline
  $ W\sub{0} $ [rad$^{-1}$] & $ 0.0227 \substack{+0.0020 \\ -0.0013} $ & $ > 0.0004 \; ( 1 \sigma ) $ \\
  $ \Delta\theta\sub{tail} $ [rad] & $ 0.172 \pm 0.006 $ & $ 0.42 \pm 0.17 $ \\
  \hline
  Ref. & V14 & R14 \\
  \hline
  \end{tabular}
  \tablebib{
  R14~\citet{2014ApJ...784...40R};
  V14~\citet[][their \mbox{1-D} model fit of the average transit profile]{2014A&A...561A...3V}.
  }
\end{table}

\subsection{Mass loss rates of the planets}
\label{s:mass_loss}

The observational constraints on the normalisation constant $ W\sub{0} $
can be used to estimate the dust mass loss rate of the evaporating planet
(i.e., excluding mass lost in gas).
Rewriting Eq.~\eqref{eq:w_0} gives
\begin{equation}
  \label{eq:mdot_obs}
  \begin{split}
    \dot{M}\sub{d}
      & \approx 0.45
        \, \Biggl( \frac{ L\sub{\star} }{ \mathrm{1~L}\sub{\odot} } \Biggr)
        \, \Biggl( \frac{ R\sub{\star} }{ \mathrm{1~R}\sub{\odot} } \Biggr)^2
        \, \Biggl( \frac{ M\sub{\star} }{ \mathrm{1~M}\sub{\odot} } \Biggr)^{-1}
        \, \Biggl( \frac{ P\sub{p} }{ \mathrm{ 1~day} } \Biggr)^{-1} \\
      & \quad \times
        \, \Biggl( \frac{ W\sub{0} }{ \mathrm{0.001~rad^{-1}} } \Biggr)
        \, \Biggl( \frac{ \bar{Q}\sub{pr} }{ 1 } \Biggr)
        \, \Biggl( \frac{ \bar{Q}\sub{ext} }{ 2 } \Biggr)^{-1}
        \; \mathrm{~M\sub{\earth}~Gyr^{-1} }.
  \end{split}
\end{equation}
For large grains,
we can approximate $ \bar{Q}\sub{pr} \approx 1 $ and $ \bar{Q}\sub{ext} \approx 2 $.
For KIC~1255b, this yields $ \dot{M}\sub{d} \approx 1.7 \pm 0.5 \mathrm{~M\sub{\earth}~Gyr^{-1} } $,
which is somewhat higher than
earlier (grain-property-dependent) estimates by
\citet{2012ApJ...752....1R}, \citet{2013MNRAS.433.2294P}, and \citet{2013ApJ...776L...6K}.
For \mbox{KOI-2700b}, we find a $ 1 \sigma $ lower limit of
$ \dot{M}\sub{d} \gtrsim 0.007 \mathrm{~M\sub{\earth}~Gyr^{-1} } $,
consistent with the estimate of \citet{2014ApJ...784...40R}.

\begin{figure}
  \includegraphics[width=\linewidth]{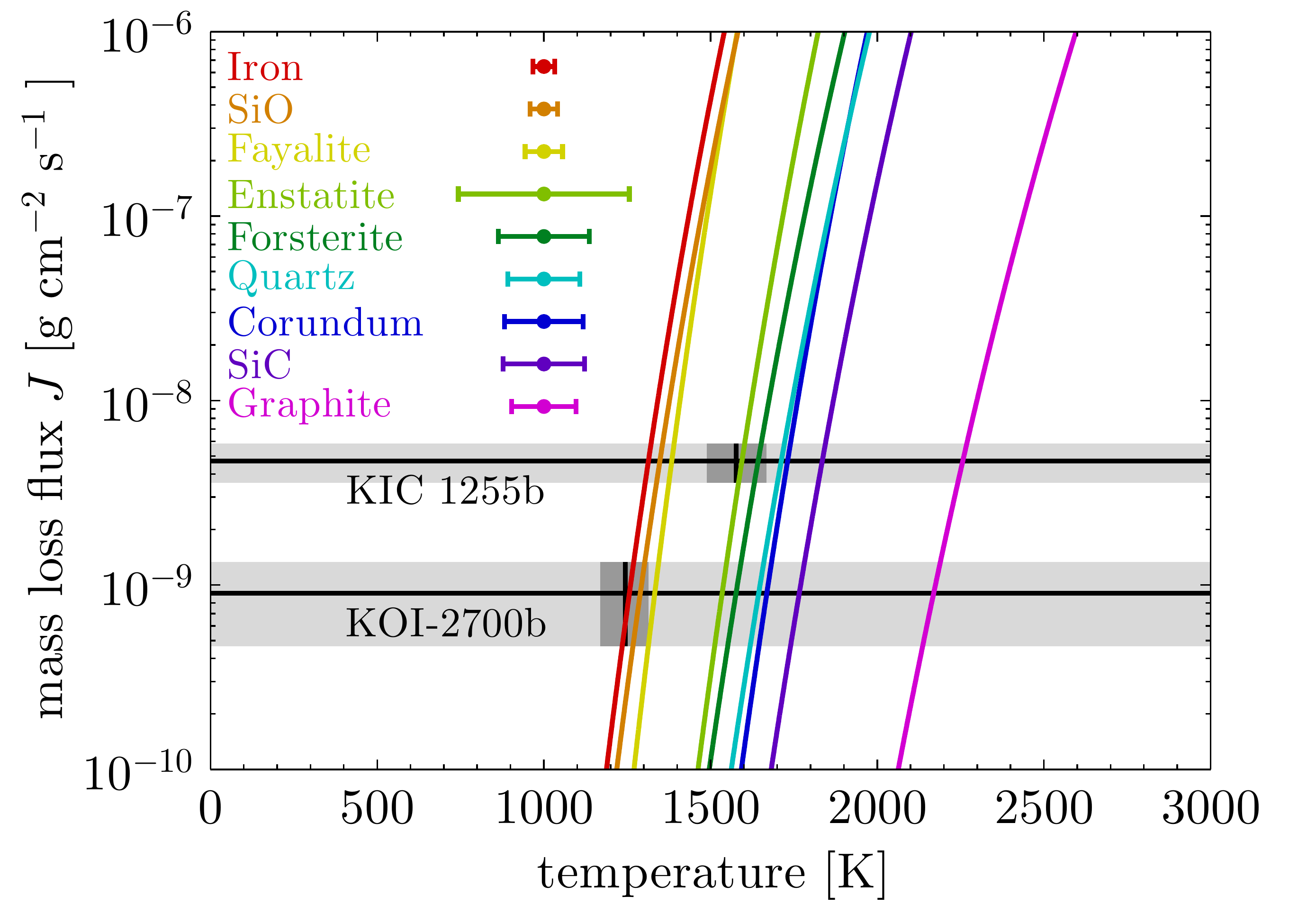}
  \caption{
  Mass loss flux as a function of temperature (Eq.~\eqref{eq:subl})
  for different materials, whose properties are listed in Table~\ref{tbl:subl}.
  Typical uncertainties (in temperature)
  are indicated by the error bars next to the legend.
  The horizontal lines mark the sublimation rates
  required to explain the observed tail lengths (Eq.~\eqref{eq:j_obs}),
  with light grey bands indicating their uncertainties.
  The black-body temperatures (Eq.~\eqref{eq:temp_bb})
  at the distance of the planets and their uncertainties
  are indicated by vertical line segments and dark grey patches.}
  \label{fig:j_temp}
\end{figure}

\subsection{Dust composition}
\label{s:composition}

The observed tail length can be used together with system parameters
and $ \bar{Q}\sub{pr} \approx 1 $ (valid for large grains)
in Eq.~\eqref{eq:tail_theory} to estimate the required orbit-averaged mass loss flux of the grains
\begin{equation}
  \label{eq:j_obs}
  \begin{split}
    \left< J \right>
      & \approx 2.1 \times 10^{-9}
        \, \Biggl( \frac{ L\sub{\star} }{ \mathrm{1~L}\sub{\odot} } \Biggr)
        \, \Biggl( \frac{ M\sub{\star} }{ \mathrm{1~M}\sub{\odot} } \Biggr)^{-1}
        \, \Biggl( \frac{ P\sub{p} }{ \mathrm{ 1~day} } \Biggr)^{-1} \\
      & \quad \times
        \, \Biggl( \frac{ \Delta\theta\sub{tail} }{ \mathrm{1~rad} } \Biggr)^{-1}
        \, \Biggl( \frac{ \bar{Q}\sub{pr} }{ 1 } \Biggr)
        \; \mathrm{ g~cm^{-2}~s^{-1} }.
  \end{split}
\end{equation}
This gives $ \left< J \right> \approx ( 4.7 \pm 1.1 ) \times 10^{-9} \mathrm{~g~cm^{-2}~s^{-1} } $ for KIC~1255b and
$ \left< J \right> \approx ( 9.0 \substack{+4.3 \\ -4.4} ) \times 10^{-10} \mathrm{~g~cm^{-2}~s^{-1} } $ for \mbox{KOI-2700b}.

In Fig.~\ref{fig:j_temp}, we compare these observational constraints on the sublimation rate
with laboratory \mbox{$ J( T ) $ curves} for different materials.
Typical uncertainties on the temperature
were estimated at $ J = 10^{-8} \mathrm{~g~cm^{-2}~s^{-1} } $ using a Monte Carlo technique
(i.e., by varying $ \mathcal{A} $ and $ \mathcal{B} $ according to their respective uncertainties,
and then numerically solving Eq.~\eqref{eq:subl} for $ T $).
We also mark the black-body temperature (Eq.~\eqref{eq:temp_bb}) at the distance of the planet for both systems.
The difference in temperature between the two systems
(greater than what can be explained by the trend of any of the material curves)
indicates that the dust in the two systems may have a different composition,
with KIC~1255b requiring a more refractory species than \mbox{KOI-2700b}.
Furthermore,
some materials reach the required sublimation rates at temperatures much lower or higher than
the typical (black-body) temperatures of the systems,
suggesting that the dust is unlikely to be composed of these materials.
However,
dust temperatures
can depart considerably from black-body temperatures
(see Fig.~\ref{fig:mat_temp}).
In addition, particles on eccentric orbits experience lower temperatures when they are farther away from the star.

\begin{figure*}
  \includegraphics[width=\linewidth]{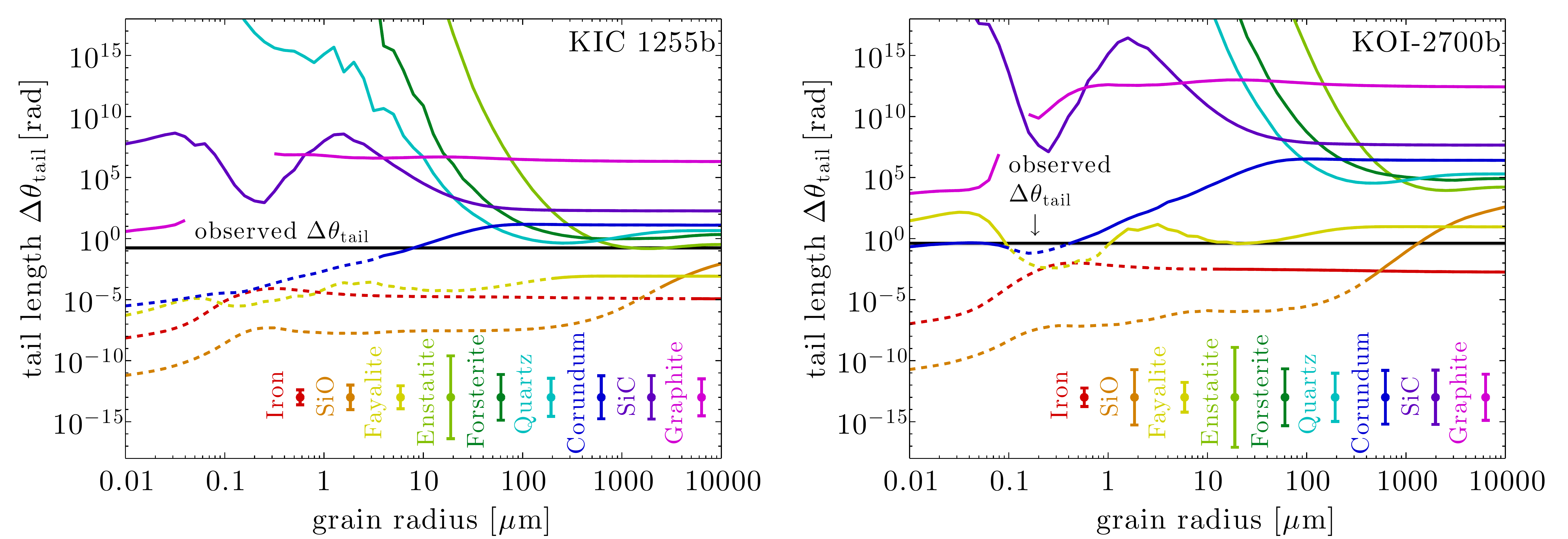}
  \caption{
  Characteristic angle of the tail's decay as a function of grain size (Eq.~\eqref{eq:tail_theory})
  for KIC~1255b and \mbox{KOI-2700b},
  with different colours for different materials.
  Dashed lines indicate that particles sublimate entirely before completing one orbit.
  Gaps (neither a solid nor a dashed line) appear where particles are unbound ($ \beta \geq 0.5 $).
  Typical uncertainties are shown by the error bars at the bottom of each panel.
  The horizontal black lines indicate the values of $ \Delta\theta\sub{tail} $ derived from the observed light curves.
  }
  \label{fig:th_s}
\end{figure*}

To investigate the effects of size-dependent grain temperatures,
we use Eq.~\eqref{eq:tail_theory} to calculate tail lengths as a function of grain size,
employing orbit-averaged sublimation rates from Eq.~\eqref{eq:subl_mean},
realistic dust temperatures from Eq.~\eqref{eq:temp_d},
and size-dependent \mbox{$ \bar{Q}\sub{pr} $ values} derived from the Mie calculations described
in Sect.~\ref{s:sigma_ext}.
The results are shown in Fig.~\ref{fig:th_s},
together with the observed tail lengths for comparison.
Typical uncertainties were estimated at $ s = 1000$~$\upmu$m using Monte Carlo simulations.
For some materials, particles of certain sizes sublimate entirely before completing one orbit
($ s\sub{0} \, / \left< \dif s / \dif t \right> < P\sub{d} $),
meaning that assumption~\ref{as:t_subl} of our derivation is violated.
In these cases, the tail length as predicted by Eq.~\eqref{eq:tail_theory}
is shown in Fig.~\ref{fig:th_s} with dashed lines
(while strictly Eq.~\eqref{eq:tail_theory} is no longer valid; also see Sect.~\ref{s:assump}).
At small grain sizes ($ s \lesssim 0.1 $~$\upmu$m),
assumption~\ref{as:qs} breaks down,
because $ \bar{Q}\sub{pr} $ then varies strongly with grain size.
At the largest grain sizes considered here ($ s \gtrsim 1000 $~$\upmu$m),
grain temperatures approach the black-body temperature,
and hence the results in Fig.~\ref{fig:th_s} converge to those of Fig.~\ref{fig:j_temp}
(but we note that such large dust grains are not very plausible; see Sect.~\ref{s:size}).

Figure~\ref{fig:th_s} displays an extremely large dynamic range in predicted tail lengths,
caused by the exponential dependence of sublimation rate on grain temperature
(see Eqs.~\eqref{eq:subl} and \eqref{eq:pres_vap2}).
For most of this range, dust tails are
probably not detectable by transit photometry.
Small tail lengths ($ \Delta\theta\sub{tail} \lesssim 0.01 $)
effectively leave only the parent planet, which makes a regular, symmetric light curve.
Large values ($ \Delta\theta\sub{tail} \gtrsim 100 $)
occur if dust grains hardly sublimate at all,
giving rise to a nearly uniform ring of dust around the star.
Such a ring would scatter and absorp the same amount of light at all phases,
and would therefore not produce a signal in the normalised light curve.

The tail length predictions of Fig.~\ref{fig:th_s} allow us to put constraints on
the composition of the dust in the tails of the two evaporating exoplanet candidates.
For KIC~1255b, we find the following:
  (1)~Iron, fayalite, and silicon monoxide
  give tail lengths that are smaller than the observed value,
  because of their low sublimation temperatures.
  (2)~The silicate minerals quartz, forsterite, and enstatite
  can produce the observed tail length, but only at very large grain sizes ($ s \gtrsim 100$~$\upmu$m).
  Smaller grains have lower temperatures (owing to their transparency in the optical),
  and therefore sublimate much slower.
  (3)~Corundum grains with sizes of around $ s \sim 10$~$\upmu$m are consistent with the observations.
  (4)~The carbonaceous materials graphite and silicon carbide
  generally give tail lengths that are much longer than the observed value,
  because of their refractory nature.
  An exception are very small ($ s \sim 0.01 $~$\upmu$m) graphite grains,
  which marginally fit the observed tail length.

The constraints for \mbox{KOI-2700b}, which is cooler, are somewhat different:
  (1)~Iron, which seems to be a good candidate in Fig.~\ref{fig:j_temp},
  yields tails that are shorter than observed, because of its higher-than-black-body temperature.
  (2)~Fayalite grains of many different sizes are consistent with the observed tail length.
  (3)~Silicon monoxide gives the observed tail length for very large grain sizes ($ s \sim 1000 $~$\upmu$m).
  (4)~The silicate minerals quartz, forsterite, and enstatite,
  as well as the carbonaceous materials graphite and silicon carbide
  all yield tails that are much longer than indicated by the observations.
  (5)~Corundum grains with sizes of $ s \lesssim 1 $~$\upmu$m give the observed tail length.

The above results are discussed in more detail in Sect.~\ref{s:implic},
where they are examined in the light of independent constraints on the typical size of the dust grains.

\section{Discussion}
\label{s:discuss}

\subsection{Validity of assumptions}
\label{s:assump}

Our semi-analytical model makes use of a number of assumptions,
listed at the beginning of Sect.~\ref{s:theory}.
In the above derivation and analysis,
we already
give arguments for some of these assumptions,
or discuss under what conditions they are valid.
Many of the assumptions, however, warrant
some further discussion,
which we provide here.
\begin{description}
\item
\textit{Steady state} (assumption~\ref{as:steady}):
It is clear that the dust tail of KIC~1255b is not in steady state,
because it experiences drastic changes in transit depth from orbit to orbit \citep{2012ApJ...752....1R}.
This variability is, in fact, an important argument for
the evaporating-planet scenario.
For \mbox{KOI-2700b}, the \textit{Kepler} data
do not have the signal-to-noise ratio required
to detect orbit-to-orbit changes in transit depth,
but a secular trend in transit depth was found \citep{2014ApJ...784...40R}.
The results we find
under the assumption of a steady-state dust tail,
using the average observed light curve,
reflect the properties of the system averaged over time.
\item
\textit{Optical depth} (assumption~\ref{as:thin}):
\citet{2012ApJ...752....1R} argue that the star-to-planet part of the dust cloud must be optically thin
or marginally so (i.e., $ \tau \lesssim 1 $)
in order for the planet to be heated sufficiently by stellar radiation to sublimate.
Indeed, \citet{2013MNRAS.433.2294P} find that
the mass loss rate of the planet goes down for high dust abundances,
and reaches a maximum at optical depths of about $ \tau \approx 0.1 $ (see their Fig.~4).
Furthermore, shielding of dust grains by each other from the stellar radiation is reduced
by the large angular diameter of the star from the vantage point of the dust cloud
($ 27 \mdegr $ for KIC~1255b; $ 19 \mdegr $ for \mbox{KOI-2700b}).
Regarding the optical depth from the star to the observer,
it is important to take into account that
the dust cloud is inclined with respect to the line of sight
(for non-zero impact parameters).
This lowers the optical depth because
it ensures that
the radial extent of the dust cloud
(which can be substantial because of
the different paths followed by grains of different sizes)
contributes to its vertical extent with respect to the line of sight.
\item
\textit{Single grain size} (assumption~\ref{as:size}):
Without knowledge of
the actual grain size distribution,
our analysis assuming a single grain size
can still be used to exclude many dust compositions.
If a given dust species
yields tail lengths that, for all plausible grain sizes,
are either always longer or always shorter than the observed value
(i.e., its curve in Fig.~\ref{fig:th_s} lies either fully above or fully below the horizontal black line),
it is not possible to create the observed tail length with a combination of
different-sized grains of this composition.
\item
\textit{Constant optical efficiencies} (assumption~\ref{as:qs}):
Figure~\ref{fig:mat_q_ext} shows for which grain sizes the extinction efficiency
remains constant.
The independence of transit depth on wavelength \citep{2014ApJ...786..100C}
indicates that this assumption is at least correct for the extinction efficiency in the case of KIC~1255b,
although the possible effects of the optical depth of the dust cloud should be investigated.
For the emission efficiency (relevant to the energy balance of the dust grains)
the assumption only becomes accurate at much larger grain sizes
(see discussion of assumption~\ref{as:dsdt} below).
\item
\textit{Survival timescale} (assumption~\ref{as:t_subl}):
We assume that particles survive against sublimation for at least one orbit.
Arguments based on the variability timescale presented by \citet{2012ApJ...752....1R} and \citet{2013MNRAS.433.2294P}
indicate that this is marginally the case for KIC~1255b.
For \mbox{KOI-2700b},
dust grains with $ \beta \lesssim 0.03 $ need to survive for longer than one dust orbit to explain the length of the tail
\citep[ignoring sublimation; see also Sect.~4.5 of][]{2014ApJ...784...40R}.
This assumption is needed to justify averaging over the orbit of the dust particle.
For low-eccentricity orbits, however,
this does not introduce a large error.
\item
\textit{Constant sublimation rate} (assumption~\ref{as:dsdt}):
Our assumption that the orbit-averaged sublimation rate
does not change as particles become smaller means that
we make a zeroth-order approximation in dust temperature.
This is only valid at very large grain sizes,
for which black-body temperatures are a good approximation.
Avoiding this simplification would require a numerical approach to modelling the dust tail.
Assumption~\ref{as:dsdt} also means that
we do not take into account
the changes in orbital eccentricity that are caused
by the changes in $ \beta $
of a shrinking dust grain.
This second point is a good approximation as long as orbital eccentricities are low,
such that the sublimation rate does not vary substantially from its periastron value.
\item
\textit{Dust temperatures} (assumption~\ref{as:temp}):
When computing dust temperatures from the energy balance Eq.~\eqref{eq:temp_d},
we assume that the contribution of latent heat due to sublimation is negligible.
This is valid for the materials and temperatures we consider \citep{1974A&A....35..197L,2014ApJ...784...40R}.
Collisional heating by stellar wind particles can also be ignored \citep{2014ApJ...784...40R}.
\item
\textit{Radiation pressure dominated dynamics} (assumption~\ref{as:dyn}):
For the dynamics of the dust grains, we only take into consideration
the gravity and radiation pressure of the star.
\citet{2014ApJ...784...40R} find that the ram pressure force due to the stellar wind
is one or two orders of magnitude lower than the radiation pressure force,
and can therefore be ignored.
Assumption~\ref{as:dyn} also requires that the initial relative velocity $ \Delta v\sub{0} $
with which the grain is launched away from the planet
is negligible compared to the radiation-pressure-induced drift.
This holds if
$ \Delta v\sub{0} \ll 2 \beta v\sub{p} $,
where $ v\sub{p} = \sqrt{ G M\sub{\star} / a\sub{p} } $ is the planet's Keplerian velocity.
Using Kepler's third law,
this can be rewritten as
\begin{equation}
  \label{eq:tail_beta_min}
  \beta \gg 0.002
  \, \Biggl( \frac{ M\sub{\star} }{ \mathrm{1~M}\sub{\odot} } \Biggr)^{-1/3}
  \, \Biggl( \frac{ P\sub{p} }{ \mathrm{ 1~day} } \Biggr)^{1/3}
  \, \Biggl( \frac{ \Delta v\sub{0} }{ \mathrm{ 1~km~s^{-1} } } \Biggr).
\end{equation}
The magnitude
of the initial relative velocity $ \Delta v\sub{0} $ is very uncertain.
It may be comparable to the escape speed of the planet,
but this depends on the precise coupling of the dust grains to the gas flow.
We do not expect $ \Delta v\sub{0} $ to be more than a few km~s$ ^{-1} $,
and therefore assumption~\ref{as:dyn} should be valid for grains with radii up to about $ s \lesssim 10 $~$\upmu$m.
Importantly, if this condition were violated,
and dust grains are launched isotropically,
some particles would end up drifting ahead of the planet.
The transit light curve resulting from such a configuration
would display a gradual ingress.
In the observed light curve, only the egress is gradual,
consistent with grain dynamics that are dominated by radiation pressure.
\end{description}

\subsection{Constraints on the grain size}
\label{s:size}

Section~\ref{s:composition} revealed that our constraints on the dust composition
are dependent on the size of the dust grains,
primarily because the grain size influences the grain temperature.
By using independent constraints on the grain size,
it is therefore possible to
make better inferences about the dust composition.
Here, we list the available constraints on the grain size
(most of which only concern KIC~1255b):
\begin{enumerate}
  \item
  \citet{2012A&A...545L...5B} find that the scattering properties of the dust trailing KIC~1255b,
  as imprinted on the light curve, are best explained by small particles (0.04~$\upmu$m~$ < s < 0.19$~$\upmu$m).
  \item
  In a similar analysis, \citet{2013A&A...557A..72B} finds that
  the pre-ingress brightening of KIC~1255b is best explain by grains of $ s \sim 0.1 $--1~$\upmu$m,
  while smaller grains ($ s \sim 0.01 $--0.1~$\upmu$m) are better at explaining its egress.
  \item
  From the independence of the transit depth of KIC~1255b on wavelength,
  \citet{2014ApJ...786..100C} derive a $ 3 \sigma $ lower limit on the grain size of
  $ s \gtrsim 0.5 $~$\upmu$m ($ s \gtrsim 0.2 $~$\upmu$m for iron grains).
  \item
  In the phase-folded short cadence light curve of KIC~1255b,
  \citet{2014ApJ...786..100C} tentatively detect a small decrement of flux
  in the egress, about 0.15 phase units after the midpoint of the transit
  (corresponding to an angular separation from the planet of
  $ \Delta\theta\sub{decr} \approx 0.15 \times 2 \pi \approx 0.94$~rad).
  If this feature is real, it may be related to the first periastron passage of the dust grains after launch.
  The periastron passage gives an enhancement in the dust density
  because the relative angular velocity between dust and planet vanishes
  when dust particles pass their periastron.
  This is illustrated by Fig.~\ref{fig:cusp_th},
\begin{figure}
  \includegraphics[width=\linewidth]{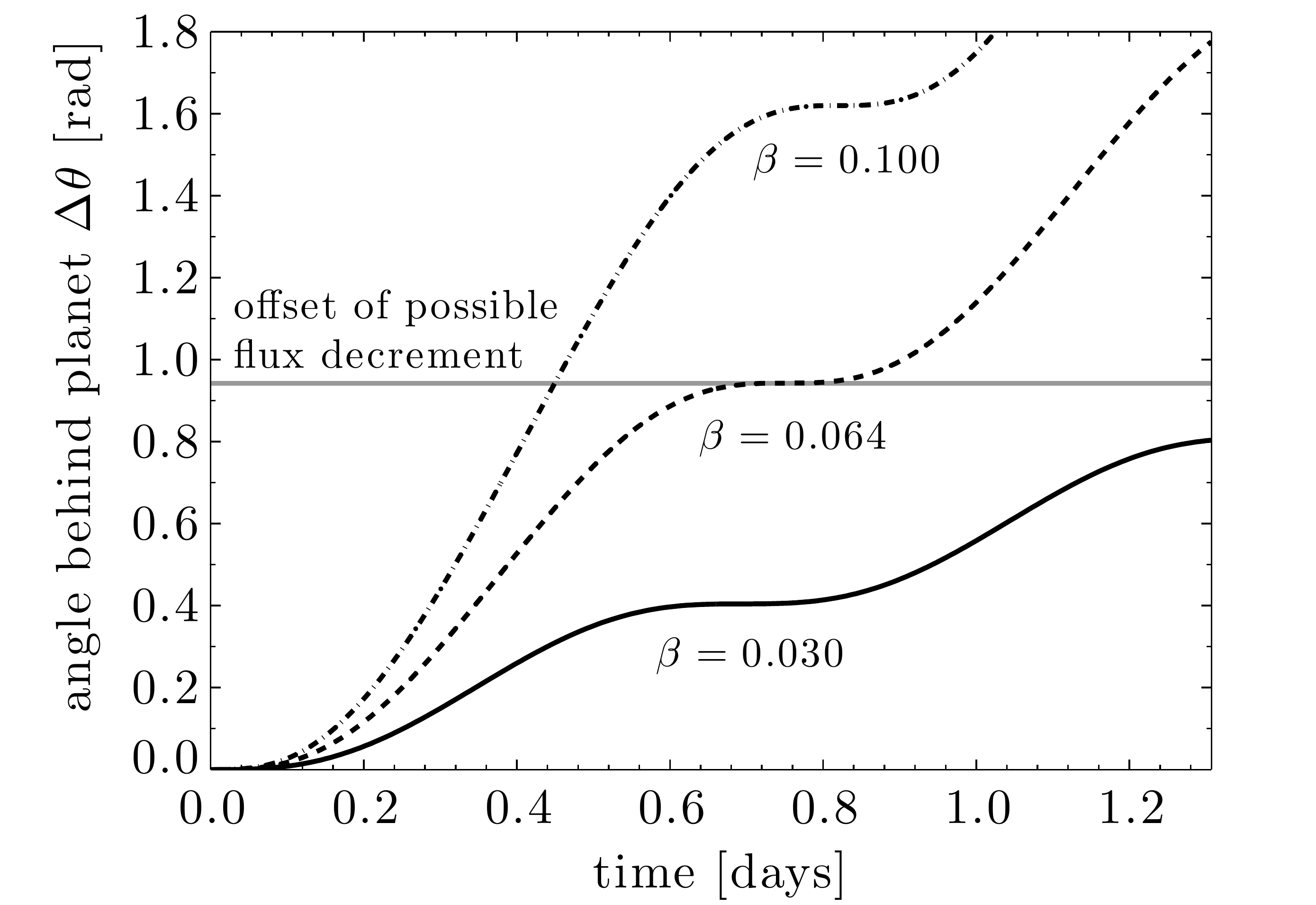}
  \caption{
  Angular separation between dust grains and the planet KIC~1255b
  as a function of time for different values of $ \beta $
  (the ratio of radiation pressure to gravitational force on a dust grain).
  The horizontal grey line indicates the position of
  the tentative flux decrement in the egress with respect to the midpoint of the transit.
  }
  \label{fig:cusp_th}
\end{figure}
  which shows the drift of dust particles with respect to the planet (ignoring sublimation),
  calculated by solving Kepler's equation
  \citep[see also Fig.~\ref{fig:sketch}, as well as Sect.~4.1 and Fig.~11 of][]{2014ApJ...784...40R}.
  The figure demonstrates that the angular offset of
  the dust density enhancement
  is very sensitive to $ \beta $,
  and hence to the size of the dust grains.
  This also implies that,
  if the flux decrement feature is real and it is caused by the periastron passage of dust grains,
  the grain size distribution must be very narrow.
  Ignoring sublimation, the relation between the \mbox{$ \beta $ ratio} of the dust grains
  and the position of the density enhancement is
  \begin{equation}
    \label{eq:cusp_th}
    \Delta\theta\sub{decr}
      = \omega\sub{syn} P\sub{d}
      = 2 \pi \frac{ 1 - \beta - ( 1 - 2 \beta )^{3/2} }{ ( 1 - 2 \beta )^{3/2} }.
  \end{equation}
  The observed angular offset of the possible flux decrement of $ \Delta\theta\sub{decr} \approx 0.94$~rad
  corresponds to $ \beta \approx 0.064 $.
  This \mbox{$ \beta $ ratio} is reached by particles of various sizes,
  depending on material type (see Fig.~\ref{fig:mat_beta}),
  but leads to a tentative upper limit of roughly $ s \lesssim 2 $~$\upmu$m,
  since larger grains always have lower \mbox{$ \beta $ ratios}.
  \item
  For very small values of $ \beta $,
  the motion of dust grains is probably dominated by the initial velocity
  with which they are launched away from the planet,
  rather than radiation-pressure-induced drift (see Eq.~\eqref{eq:tail_beta_min}).
  Large grains may therefore result in
  a more symmetric distribution of dust around the planet,
  contrary to what is observed.
  For this reason, we deem dust grains with sizes of $ s \gtrsim 100 $~$\upmu$m unlikely.
  \item
  \citet{2013MNRAS.433.2294P}
  assume $ s = 1 $~$\upmu$m
  for their atmospheric outflow model,
  and find that grains up to this size
  couple sufficiently to the gas to be lifted out of the planetary atmosphere by a thermal wind.
  Although the resulting upper limit on grain size is model-dependent
  (it varies with planet mass, for example),
  this is an additional argument against very large grain sizes.
\end{enumerate}
Combined, the above pieces of evidence point to
grain sizes of about \mbox{$ s \sim 0.1 $--$ 1 $~$\upmu$m} for KIC~1255b,
although some of the clues are contradictory.
For both objects, there are arguments to exclude very large dust grains ($ s \gtrsim 100 $~$\upmu$m).
Future research should examine whether
a uniform grain size
is a good approximation.
It is also important to
determine the optical depth of the dust clouds,
and investigate what effects it may have
on the shape of the pre-ingress brightening and on the wavelength dependence of the transit depth.

\subsection{The composition of the dust and the planets}
\label{s:implic}

Applying the above grain size constraints to the results from Sect.~\ref{s:composition}
allows us to exclude some of the solutions seen in Fig.~\ref{fig:th_s}.
Specifically, the silicate minerals that reproduce the observed tail length of KIC~1255b
at very large grain sizes now seem unlikely.
The same holds for SiO in the case of \mbox{KOI-2700b}.

One of the remaining candidates for the dust material is corundum (Al$_2$O$_3$).
It yields the observed tail lengths of both KIC~1255b and \mbox{KOI-2700b}
at plausible grain sizes.
Because of the relatively low cosmic abundance of aluminium,
it may seem unlikely that
a minor species such as corundum
accounts for all the dust
in the tails of the evaporating planets.
However, as already noted by \citet{2012ApJ...752....1R},
high Al$_2$O$_3$ abundances
can be created by a distillation of the planetary surface.
This mechanism was proposed by \citet{2011Icar..213....1L}
in a study of the lava ocean
which is thought to dominate the dayside surface of the hot super-Earth \mbox{CoRoT-7b}.
Starting with a silicate composition,
more volatile elements such as Si, Fe, and Mg evaporate preferentially from the lava ocean,
and, if these components leave the system permanently,
the ocean over time consists of increasingly refractory species.
After about 1.5~Gyr of steady evaporation,
the ocean reaches a stable composition of $ 13 \%$ CaO and $ 87 \% $ Al$_2$O$_3$.
A similar distillation scenario for KIC~1255b and \mbox{KOI-2700b}
may explain the prevalence of corundum grains.

Previous work on the evaporating exoplanet candidates
proposed pyroxene ([Mg,Fe]SiO$_3$)
as the main constituent of the dust in the tails,
and (in the case of KIC~1255b) rejected olivine ([Mg,Fe]$_2$SiO$_4$) as too volatile
\citep{2012ApJ...752....1R,2014ApJ...784...40R}.
These suggestions are based on scaling the sublimation time scales
found by \citet{2002Icar..159..529K} for cometary dust grains in the vicinity of the Sun
to the exoplanetary systems.
Here, we show that dust grains made of enstatite (the \mbox{Mg-rich} end-member of pyroxene)
survive for too long to explain the observed tail lengths.\footnote{
We only considered crystalline minerals,
because amorphous forms are expected to transition to crystalline
on timescales much shorter than the sublimation timescale \citep[see Sect.~4.2.3 of][]{2002Icar..159..529K}.}
Regarding olivine, the \mbox{Mg-rich} end-member forsterite behaves similar to enstatite,
while the \mbox{Fe-rich} end-member fayalite gives much shorter tails.
For \mbox{KOI-2700b}, the observed tail length is consistent with dust grains composed of fayalite,
while for KIC~1255b fayalite is too volatile.
Since these two end-members of olivine give tail lengths on opposite sides of the observed value for KIC~1255b,
it is conceivable that an intermediate form (iron-rich, but with some magnesium) yields the right length.

The list of dust materials we tested is far from exhaustive.
In addition, we only considered dust grains of a single, pure composition.
This type of investigation only allows one to test whether a dust species
is consistent with the observations or not,
and there may be more dust species, different from the ones we found, that can also explain the data.
Perhaps the most stringent constraint resulting from this work is that
pure iron and carbonaceous dust compositions are not favoured.

Broadly speaking, the composition of the dust ejected by evaporating planets
reflects that of the parent planet.
Hence, pure iron and carbonaceous compositions are not favoured
for the planets KIC~1255b and \mbox{KOI-2700b}.
However, the relation between dust and planet composition
may be complicated by processes such as
preferential condensation of certain species and fractionation of a lava ocean \citep{2011Icar..213....1L}.
Therefore, to gain more insight into the planetary composition,
a thorough investigation of the link between planet and dust composition is required.
Roughly, this would entail the following modelling steps:
(1)~Given a planetary composition, temperature, and gravity,
calculate the composition of the atmosphere resulting from evaporation
\citep[e.g.,][]{2009ApJ...703L.113S,2011ApJ...742L..19M,2012ApJ...755...41S}.
(2)~Determine the dynamical structure of the atmospheric outflow
\citep[see][]{2013MNRAS.433.2294P}.
(3)~Calculate which dust species would condensate in this outflow
\citep[as has been done for stellar outflows; e.g.,][]{2010LNP...815...61G}.\footnote{
This assumes that the outflow is loaded with dust grains as a result of condensation from the gas phase.
Another possible origin of the dust is explosive vulcanism
\citep[see Sect.~4.2 of][and references therein]{2012ApJ...752....1R}.}
By doing this for different potential planetary compositions,
and as a function of temperature and gravity,
it may be possible to
use the dust composition as a probe for
the composition of the planet.

\section{Conclusions}
\label{s:conclusions}

This paper describes the decay of cross-section in dusty tails trailing evaporating planets,
such as KIC~1255b and \mbox{KOI-2700b}.
The analytical expression we derive (Eq.~\eqref{eq:tail_decay_total})
can be used to model transit light curves.
Specifically, it provides a physical interpretation for
two properties of a dust tail.
The density of the tail (found from the transit depth) is related to the mass loss rate of the planet.
The tail length (which can be derived from the duration of the transit) is determined by the sublimation rate of the dust in the tail.
This sublimation rate depends sensitively on the optical and thermodynamical properties
of the material that the dust grains in the tail are made of.
Therefore, given accurate laboratory measurements of these properties,
the tail length can be used to constrain the composition of the dust.

\begin{table*}[!t]
  \centering
  \caption{Summary of the dust composition constraints}
  \label{tbl:sum}
  \begin{tabular}{lcc}
  \hline
  Dust species & KIC~1255b & KOI-2700b \\
  \hline
  Iron (Fe) & too volatile & dust temperature too high \\
  Silicon monoxide (SiO) & too volatile & required grain size implausibly large \\
  Cryst. fayalite (Fe$_2$SiO$_4$) & too volatile & consistent with observed tail length \\
  Cryst. enstatite (MgSiO$_3$) & required grain size implausibly large & too refractory \\
  Cryst. forsterite (Mg$_2$SiO$_4$) & required grain size implausibly large & too refractory \\
  Quartz (SiO$_2$) & required grain size implausibly large & too refractory \\
  Corundum (Al$_2$O$_3$) & consistent with observed tail length & consistent with observed tail length \\
  Silicon carbide (SiC) & too refractory & too refractory \\
  Graphite (C) & too refractory & too refractory \\
  \hline
  \end{tabular}
\end{table*}

The constraints we find for the dust composition of the two evaporating exoplanet candidates
are summarised in Table~\ref{tbl:sum}.
Our analysis
has lead to the following conclusions about these systems:
\begin{enumerate}
  \item
  The mass loss rate in dust of KIC~1255b is approximately
  $ \dot{M}\sub{d} \approx 1.7 \pm 0.5 \mathrm{~M\sub{\earth}~Gyr^{-1} } $.
  For \mbox{KOI-2700b}, we find a $ 1 \sigma $ lower limit of
  $ \dot{M}\sub{d} \gtrsim 0.007 \mathrm{~M\sub{\earth}~Gyr^{-1} } $.
  \item
  Dust grains composed of corundum can explain the tail lengths of both candidates.
  \item
  The tail length of \mbox{KOI-2700b} is also consistent with a fayalite composition.
  A composition of iron-rich silicate minerals may also work for KIC~1255b.
  \item
  Dust grains made of pure iron, graphite, or silicon carbide are not favoured for both objects.
\end{enumerate}

\begin{acknowledgements}
We thank H.-P. Gail for sharing refractive index data of SiO.
We also appreciate the constructive comments of an anonymous referee.
\end{acknowledgements}

\bibliographystyle{aa}
\bibliography{bib_ads.bib}

\end{document}